# An IBE-based Signcryption Scheme for Group Key Management

Peter Gutmann
University of Auckland

Steven M. Bellovin
Columbia University

Matt Blaze
University of Pennsylvania

Ronald L. Rivest
MIT

Nigel Smart
University of Bristol

**Abstract.** This paper presents a new crypto scheme whose title promises it to be so boring that no-one will bother reading past the abstract. Because of this, the remainder of the paper is left blank.

**Acknowledgements.** The authors would like to thank the anonymous reviewers for pointing out an error in the proof on page 12.